# Transmutation of singularities and zeros in graded index optical instruments: a methodology for designing practical devices

I. R. Hooper[*] and T. G. Philbin

*Electromagnetic and Acoustic Materials Group, Department of Physics and Astronomy, University of Exeter, Exeter, EX4 4QL, UK*
[*]*i.r.hooper@exeter.ac.uk*

**Abstract:** We describe a design methodology for modifying the refractive index profile of graded-index optical instruments that incorporate singularities or zeros in their refractive index. The process maintains the device performance whilst resulting in graded profiles that are all-dielectric, do not require materials with unrealistic values, and that are impedance matched to the bounding medium. This is achieved by transmuting the singularities (or zeros) using the formalism of transformation optics, but with an additional boundary condition requiring the gradient of the co-ordinate transformation be continuous. This additional boundary condition ensures that the device is impedance matched to the bounding medium when the spatially varying permittivity and permeability profiles are scaled to realizable values. We demonstrate the method in some detail for an Eaton lens, before describing the profiles for an "invisible disc" and "multipole" lenses.

**1. Introduction**

Since the seminal works by Leonhardt [1] and Pendry *et al.* [2] in 2006 a large body of theoretical work has been undertaken developing the tools and framework of Transformation Optics (TO), a full review of which is beyond the scope of this work but can be found elsewhere [3-5]. The basic principles, however, are rather straightforward. The ray trajectories (and fields) of an electromagnetic wave are tied to their co-ordinate system, and any deformation of that co-ordinate system will result in a corresponding deformation of the ray trajectories. Since Maxwell's equations have the same form in any co-ordinate system with only the permittivity and permeability being different, any co-ordinate transformation is equivalent to a new medium with, in general, spatially varying permittivity and permeability profiles. Therefore, if a particular ray trajectory is desired TO will supply the spatially varying permittivity and permeability profile required to produce it.

Though the advent of TO promised a wealth of novel refractive devices, in reality there have been relatively few designs, and even fewer of these have been practically realised.

Designs have included various types of cloaking schemes [6-10], energy harvesters [11], high directivity antennas [12,13], concentrators [14] and adaptations of graded index lens designs such as the Eaton, Luneburg and Maxwell fish-eye lenses [15-21]. The lack of practical progress can be attributed to a fundamental limitation in the TO recipe: the resulting permittivity and permeability profiles frequently require extreme values that are not available in naturally occurring media and are difficult to achieve even with metamaterials [22-24]. Moreover, TO-developed designs require the permittivity and permeability tensors to be equal (in other words the materials prescribed by TO are impedance-matched) which is an even greater challenge from a materials perspective. It is well known that naturally occurring magnetic materials may only exhibit low-loss broadband magnetic responses at low frequencies (<100MHz) [25,26], and that enabling a magnetic response at any other frequency (whether through the use of ferromagnets or metamaterials) requires a resonant response [27]. This requirement limits the bandwidth of a device and also results in losses, which may not be tolerable depending on the application. The permittivity and permeability profiles are also, in general, anisotropic and, whilst certain levels of anisotropy can be incorporated (especially in metamaterials), the experimental design becomes even more demanding. Schemes to overcome these limitations have led to adaptations of the TO recipe such as the recent work on quasi-conformal mapping which, for example, has been used to design carpet cloaks [8,9,28,29], and a relaxation of the impedance-matching requirement of TO for single polarisation operation [30].

It is therefore clear that, if one requires broadband operation, non-resonant materials must be used. This limits the palette of available materials to those without a magnetic response (or, more generally, a diamagnetic response) that cannot be impedance-matched, requiring an adaptation of the TO recipe. One such adaptation involves a simple rescaling of the permittivity and permeability tensors, and has been used in some experimental demonstrations of TO designed devices [31,32]. However, this method has two significant downsides: it limits device operation to a single polarisation, and reflections arise from the boundary of the device since it is no longer impedance-matched. In the next section we will show that, though the limitation of single polarisation operation cannot be overcome, the reflections arising from the boundary of the device can be removed through a simple modification of the co-ordinate transformation in the same manner as Cai *et al*. [6], who used a higher-order co-ordinate transformation in the design of a cloak resulting in impedance matching at the boundary of the device. However, whilst their cloak design resulted in an all-dielectric profile that was impedance matched at the boundary, certain permittivity tensor components were less than unity, requiring a resonant response with associated losses and narrow-band character. In the spirit of practicality we would like to design devices that can both be genuinely realizable and broadband. As such we will use a similar design process to that of Cai *et al*., but in devices where all resulting material properties are in principle achievable for low-loss, broadband operation. In particular we will demonstrate the efficacy of the design process in the transmutation of singularities and zeros in the refractive index profiles of graded index optical devices. We will use the transmuted cylindrical (2D) Eaton lens as the example in a detailed description of the method, before describing other transmuted devices including the "invisible disc" and "multipole" lenses.

## 2. Design process

The Eaton lens was first described in a paper published in 1952 [16] and is a spherically symmetric graded index lens that acts as an omnidirectional retro-reflector, with a refractive index profile given by

$$n(r) = \sqrt{\frac{2}{r} - 1} \text{ if } r \leq 1, \text{ and } n(r) = 1 \text{ if } r > 1 \tag{1}$$

where $r$ is the radial co-ordinate and we assume a unit radius. The ray trajectories in a single plane for light incident upon an Eaton lens are shown in Fig. 1(a). Whilst the isotropic refractive index profile defined by eqn. 1 is rather simple, it is clear that as $r \to 0$ the refractive index becomes singular, preventing an Eaton lens from being fabricated. Tyc *et al.* [33] demonstrated that a point singularity could be removed by expanding space around it, provided the divergence is not too steep, with the isotropic refractive index profile being transformed into an anisotropic impedance-matched permittivity (permeability) profile. Recently, Ma *et al.* [32] demonstrated a transmuted cylindrical Eaton lens with the permittivity and permeability profile achieved through the use of split ring resonator based metamaterials. However, due to the resonant nature of the metamaterial elements this was a narrow band device and thus of limited practical use.

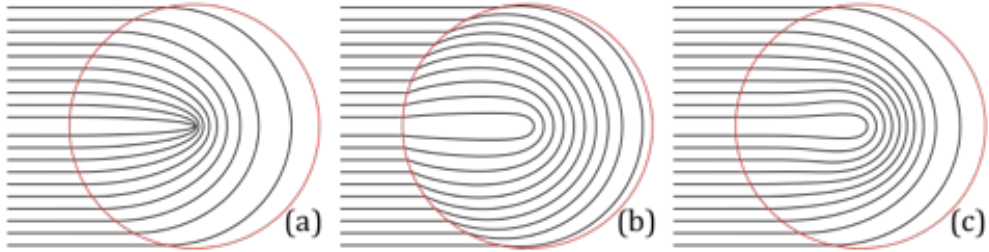

Figure 1. Ray trajectories in a plane through a) a spherical Eaton lens, b) an impedance-matched transmuted spherical Eaton lens using the co-ordinate transformation in eqn. 2, c) a transmuted cylindrical Eaton lens with non-magnetic materials using the co-ordinate transformation in eqn. 10.

The transmutation process may be applied to either the entire radius of the lens, or to only the region around the singularity [34]. Here we will consider the case where the entire lens is transmuted since this results in less extreme values of the resulting permittivity and permeability profiles, but the mathematics will be written for the general case. The co-ordinate transformation for the expansion of space around the singularity is given by:

$$R(r) = \frac{f(r)}{N(b)} \text{ if } r \leq b, \text{ and } R(r) = r \text{ if } r > b \tag{2}$$

where

$$f(r) = \frac{\int_0^r n(r')dr'}{\int_0^1 n(r')dr'} \tag{3}$$

$R$ is the transformed radial co-ordinate, $r$ is the original radial co-ordinate, and b is the radius of the transmuted region. $N(b) = f(r)|_{r=b}$ and is a normalisation to ensure that the co-ordinate transformation is continuous at the boundary of the transmuted and non-transmuted regions.

It is a simple matter to calculate the material properties arising from this expansion of space [3,22,34]. Here we present only the final result using the notation used in [3]:

$$\varepsilon_k^i = \mu_k^i = n \operatorname{diag}\left(\frac{r^2}{R^2}\frac{dR}{dr},\ \frac{dr}{dR},\ \frac{dr}{dR}\right) \tag{4}$$

where $\varepsilon_k^i$ and $\mu_k^i$ are the permittivity and permeability tensors in spherical co-ordinates. The resulting material properties are shown in Fig. 2(a), demonstrating that the singularity has been removed at the cost of substantial anisotropy. The ray trajectories resulting from the material properties shown in Fig. 2(a) are shown in Fig. 1(b), demonstrating that the retro-reflecting nature of the device has been retained.

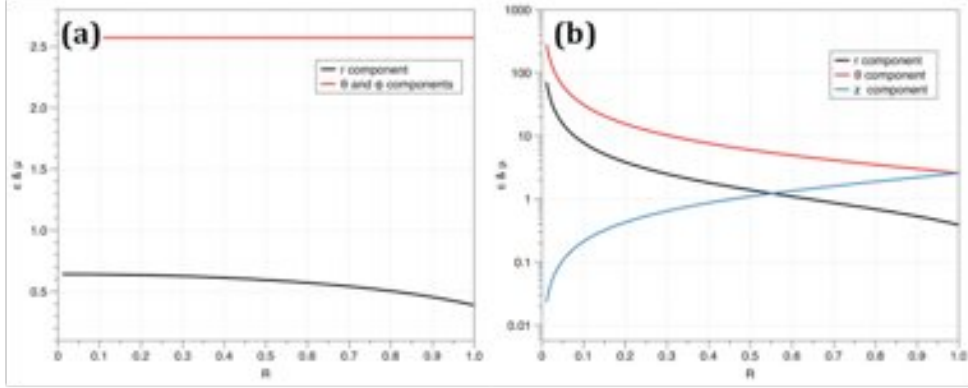

Figure 2. (a) The permittivity and permeability as a function of the radius for the transmuted spherical Eaton lens. (b) The permittivity and permeability as a function of the radius obtained using the same transmutation process, but in cylindrical co-ordinates (note the log scale).

For a cylindrical Eaton lens the point singularity becomes a line singularity, and the equivalent to equation 4 is:

$$\varepsilon_k^i = \mu_k^i = n \operatorname{diag}\left(\frac{r}{R}\frac{dR}{dr},\ \frac{R}{r}\frac{dr}{dR},\ \frac{r}{R}\frac{dr}{dR}\right) \tag{5}$$

where the three components are the permittivity & permeability in the r, θ and z directions respectively. The resulting material properties are shown in Fig. 2(b) (note the log scale), and it is clear that the singularity is no longer transmuted in the individual ε & μ components. If we now determine the refractive index tensor by combining the relevant permittivity and permeability tensor components, we obtain

$$n' = \operatorname{diag}(\sqrt{\varepsilon_\theta \varepsilon_z},\ \sqrt{\varepsilon_r \varepsilon_z},\ \sqrt{\varepsilon_r \varepsilon_\theta}) = n \operatorname{diag}\left(\frac{dr}{dR},\ \frac{r}{R},\ 1\right) \tag{6}$$

where $n'$ is the refractive index profile corresponding to the transmuted geometry, and we have described the refractive index tensor components in terms of the permittivity components with the understanding that they are interchangeable with the permeabilities. The resulting refractive index profile is shown in Fig. 3, and it is clear that the singularity has been transmuted in both the r & θ directions but not in the z direction (unlike in the spherical case where the singularity was removed in all components). By limiting ourselves to propagation in the plane orthogonal to the line singularity, and therefore negating any influence of the z component of the refractive index on the ray trajectories, we would expect the cylindrical device to exhibit the same characteristics as the spherically symmetric device. Indeed it is an important point to note that line singularities and line zeros are only ever transmuted in the resulting refractive index profile (and can never be transmuted in the individual permittivity and permeability tensor components). We also note that, whilst point singularities can always be transmuted (as long as the refractive index profile does not increase more rapidly than $1/r^p$), and line singularities can be transmuted for linearly polarised light when the electric vector is parallel of perpendicular to the singularity, higher order singularities and zeros, such as those found in the well-known cloaking schemes, can never be transmuted.

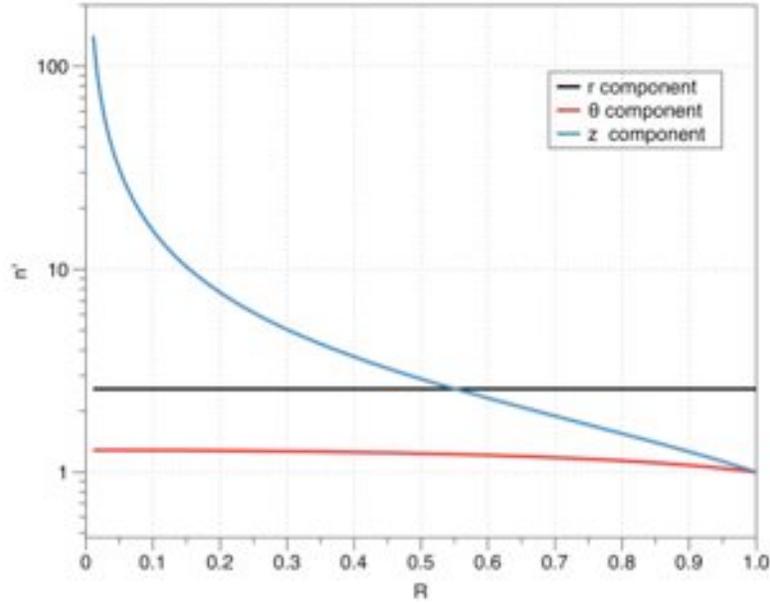

Figure 3. The components of the refractive index tensor as a function of the radial distance for the transmuted cylindrical Eaton lens obtained using eqn. 6.

Since we cannot transmute the singularities in the individual permittivity and permeability components we can no longer satisfy the impedance matching ($\varepsilon = \mu$) condition, and this results in two choices for the calculation of the refractive index that correspond to the two linear polarizations. These are:

$$\text{Electric field in the } \{r, \varphi\} \text{ plane: } n_r = \sqrt{\varepsilon_\theta \mu_z} \ \& \ n_\theta = \sqrt{\varepsilon_r \mu_z} \qquad (7)$$

$$\text{Magnetic field in the } \{r, \varphi\} \text{ plane: } n_r = \sqrt{\mu_\theta \varepsilon_z} \ \& \ n_\theta = \sqrt{\mu_r \varepsilon_z} \qquad (8)$$

and we can choose any combination of the permittivity and permeability components that results in the refractive index profiles shown in Fig. 3.

Having described the transmutation of singularities and zeros in the profiles of graded index devices, we must return to the question of practicality, and in particular we must consider the materials available for fabricating such a device. As mentioned previously, if one wishes to design a device that can operate over a broad frequency band and exhibits low losses, one is restricted to the use of materials that exhibit little or no resonant response. Also, since a large degree of anisotropy is required in the transmuted device material properties, we are likely limited to the use of metamaterials. Whilst non-resonant metallic metamaterials may exhibit very high dielectric responses they are unable to provide permeabilities of greater than unity, though they will often exhibit diamagnetic responses ($0 \leq \mu \leq 1$) [35-37]. Also, the refractive index of such metamaterials must also always be greater than unity. We can summarise these limitations as follows:

$$\begin{aligned} \varepsilon &> 1 \\ 0 \leq \mu &< 1 \\ n &\geq 1 \end{aligned} \qquad (9)$$

By inspection of eqn. 8, it is clear that we cannot satisfy the conditions of eqn. 9 for light polarised with its magnetic field in the $\{r, \varphi\}$ plane, and a broadband transmuted cylindrical Eaton lens design is not possible for this polarisation. However, from eqn. 7, for light polarised with its *electric* field in the $\{r, \varphi\}$ plane, it appears that the conditions of eqn. 9 can be completely satisfied. Here, for simplicity, we will limit ourselves to materials with a purely dielectric response ($\mu = 1$) as opposed to the more likely diamagnetic response of the non-resonant metallic metamaterials discussed above. We can then determine the permittivity profile of the device using eqn. 7, and the resulting material properties are shown in Fig. 4.

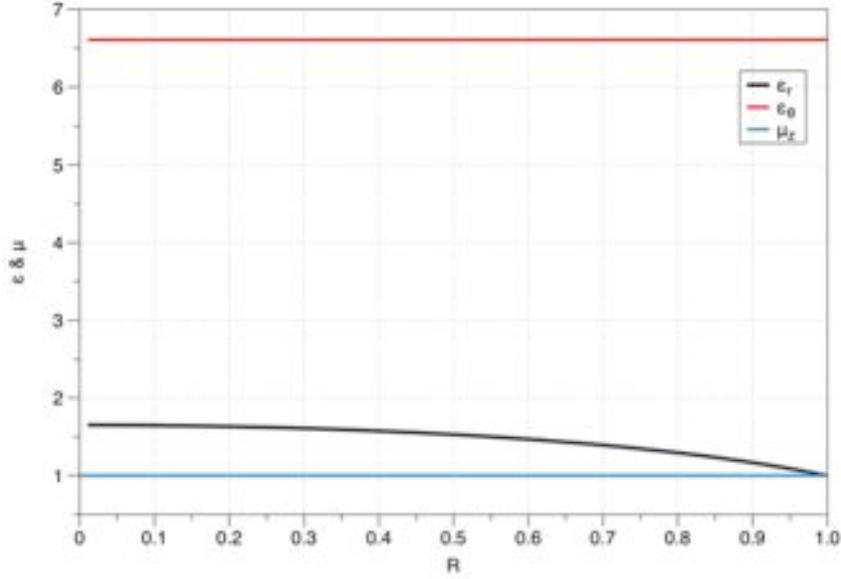

Figure 4. The relevant permittivity and permeability components as a function of the radial co-ordinate of a transmuted cylindrical Eaton lens for light polarised with the electric field in the $\{r, \varphi\}$ plane.

Unfortunately the process of scaling the permittivity and permeability profiles whilst maintaining the refractive index profile gives rise to a further drawback. Even though the ray trajectories for light incident upon a device described by the permittivity and permeability profiles shown in Fig. 4 are identical to those of the non-scaled version there is one important difference: the scaled, and therefore no longer impedance-matched, device will exhibit reflections at the boundary between the non-transmuted and transmuted regions. Indeed these reflections are found to have a significant degradation on the performance of the transmuted cylindrical Eaton lens as can be seen in Fig. 5(a) in which a commercially available finite element modeling package (COMSOL Multiphysics) has been used to model a narrow Gaussian beam incident upon the device. Strong reflections are evident from the boundary of the lens resulting in only approximately 60% of the incident beam being retro-reflected as opposed to the 100% from the equivalent impedance-matched device. The remaining light is scattered in all directions.

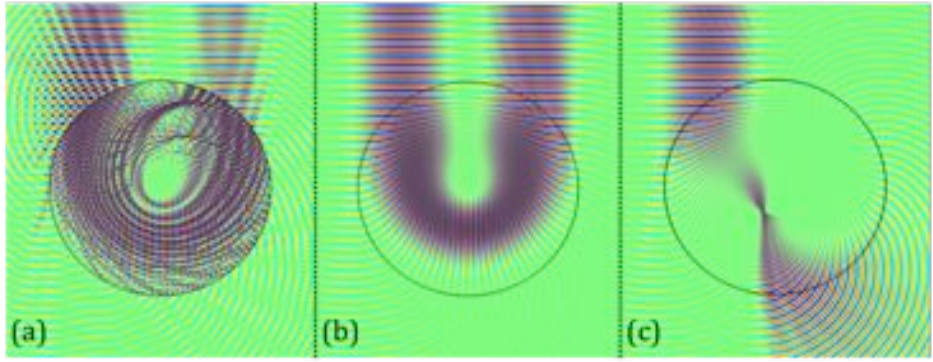

Fig 5. Full wave numerical modeling (instantaneous total electric field) of transmuted cylindrical Eaton lenses. a) a scaled transmuted Eaton lens consisting of non-magnetic materials (in-plane polarisation) using the co-ordinate transformation given by eqn. 2 b) a transmuted Eaton lens consisting of non-magnetic materials using the modified co-ordinate transformation given by eqn. 8 (in-plane polarisation). c) The same as b), but for the out-of-plane polarisation.

The reflections arise due to the discontinuity in $\varepsilon$ and $\mu$ at the boundary of the device (Fig. 4), and upon inspection of eqn. 6 it is clear that the only term that can lead to this discontinuity is the gradient $dR/dr$. Indeed, whilst the $N(b)$ term in the original co-ordinate transformation (eqn. 2) was introduced to ensure that the transformation was continuous at the boundary, there was no such constraint on the gradient. It becomes obvious, therefore, that if we add a new boundary condition requiring the gradient of the transformation be continuous at the boundary of the device, we should eliminate any reflections even when using non-impedance-matched materials. In essence we will be forcing the material properties of the transformed medium to be matched to the bounding medium. Of course, we must also note that, if we wish to achieve a device that exhibits close to zero scattering, the size of the device must be sufficiently large relative to the wavelength such that the spatially varying refractive index does not give rise to additional scattering.

There are many ways in which this new boundary condition can be imposed. Here we simply add a polynomial to the original co-ordinate transformation to demonstrate the principle:

$$R(r) = \frac{f(r)}{N(b)} + \alpha r^m (r - b) \qquad (10)$$

where m is any positive number greater than a certain limit that depends upon the original refractive index profile (or else the singularity is no longer transmuted), and $\alpha$ is a constant yet to be determined. This functional form ensures that when $r = 0$ and $r = b$ the co-ordinate transformation is unaltered from the original transformation and remains continuous. We now add the new boundary condition requiring that the gradient of the transformation be continuous across the boundary between the transmuted and non-transmuted regions:

$$\left.\frac{dR}{dr}\right|_{trans, r=b} = \left.\frac{dR}{dr}\right|_{non-trans, r=b} \qquad (11)$$

For the simple case of a fully transmuted device bounded by free space $dR/dr|_{r=b} = 1$. Using this condition, and a choice for the value of m, it is a simple matter to determine the value of alpha, and to subsequently determine the material properties corresponding to this new co-ordinate transformation in the same way as previously. The material properties for such an adapted fully transmuted cylindrical Eaton lens as a function of the radius for $m = 1$ and for non-magnetic materials are shown in Fig. 6, with the original and modified co-ordinate transformations shown inset. Whilst the choice of value for $m$ is arbitrary (the device will still result in the same behavior as the original un-transformed Eaton lens) it has a significant impact upon the resultant material properties. If $m$ is large the co-ordinate transformations more closely resemble each other with the gradient matching "squeezed" towards the outer radius of the device. This results in a lower required permittivity, but a more rapid rate of change of permittivity as a function of the radial co-ordinate close to the boundary of the device. If $m$ is smaller the gradient matching occurs over a larger radial distance and the changes to the spatially varying permittivity are "squeezed" towards the center of the device. A larger range of permittivity is required in this latter case. The choice of $m$ therefore depends upon the size of the device and the range of material properties available. More generally, the way in which the additional boundary condition is implemented will have a significant impact upon the resulting material properties.

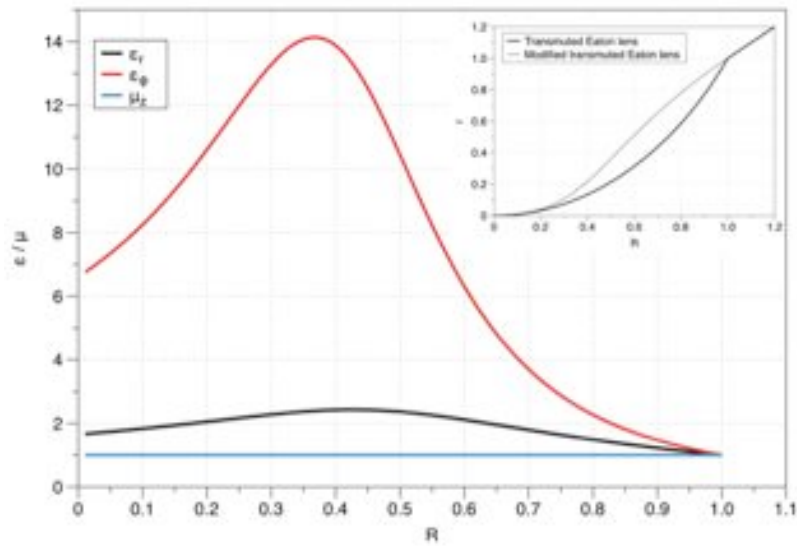

Fig 6. The permittivity and permeability profiles of the modified transmuted Eaton lens (for light polarised with its electric field in the $\{r, \varphi\}$ plane. Inset: the co-ordinate transformations of the transmuted and modified transmuted Eaton lenses.

To demonstrate the efficacy of the device the ray trajectories for the design are shown in Fig. 1(c), with full wave numerical modeling of the device for an incident Gaussian beam polarised with its electric field in the $\{r, \varphi\}$ plane shown in Fig. 5(b) (for comparison the case for light polarised with its magnetic field in the $\{r, \varphi\}$ plane is shown in Fig. 5(c)). The lack of refraction at the boundary of the device in Fig. 1(c) is clear evidence of the matching to free space that the adapted co-ordinate transformation has accomplished, whilst the full wave simulations demonstrate that the reflections that limited the performance of the un-modified transmuted Eaton lens have been removed.

In summary, the methodology we have described in this section allows for the design of practical devices in which singularities or zeros in the profile of graded index devices have been removed. This is achieved through the transmutation process, but with the addition of a boundary condition requiring the gradient of the co-ordinate transformation be continuous at the boundary of the device. A subsequent scaling of the permittivity and permeability profiles allows for all-dielectric designs that could in principle be fabricated using current metamaterial designs. We also note that the method can be applied to spherical systems as well as the cylindrical system considered here with the simple addition of a spatially varying φ component of the permittivity tensor.

## 3. Other transmuted devices

The same process as described for the cylindrical Eaton lens can be performed on other graded index devices that incorporate singularities or zeros in their refractive index. The invisible disc is one such device, and is named as such because any rays impinging upon the device exit along the exact same trajectory as they would have were the device not present, but having undergone a single loop within the device (see Fig. 7). The refractive index profile of the invisible disc is given by

$$n(r) = \left(Q - \frac{1}{3Q}\right)^2 \tag{12}$$

where

$$Q = \sqrt[3]{-\frac{1}{r} + \sqrt{\frac{1}{r^2} + \frac{1}{27}}} \tag{13}$$

and r is the radial co-ordinate.

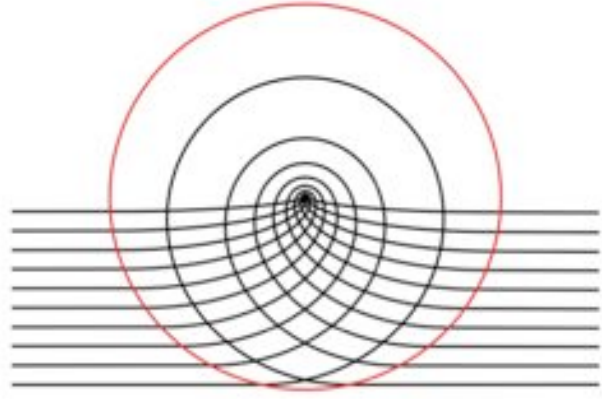

Figure 7. Ray trajectories for an "invisible disc". Any ray impinging upon the device exits as if the device were not present.

Though the details of the mathematics is somewhat more involved than in the case of the Eaton lens due to the more complex refractive index profile, the exact same process can be applied, resulting in the permittivity profile shown in Fig. 8 (left) for a value of $m = 1$. By simple comparison of Fig. 8 with the equivalent plot for the Eaton lens shown in Fig. 6 it is immediately obvious that the required material properties for the invisible disc are more extreme, both in magnitude and in anisotropy, than are those for the Eaton lens. This is a direct consequence of the increased gradient of the refractive index as a function of the radial co-ordinate of the original invisible disc profile, which is required to bend the rays sufficiently to complete a full loop within the device as opposed to the half loop of the Eaton lens.

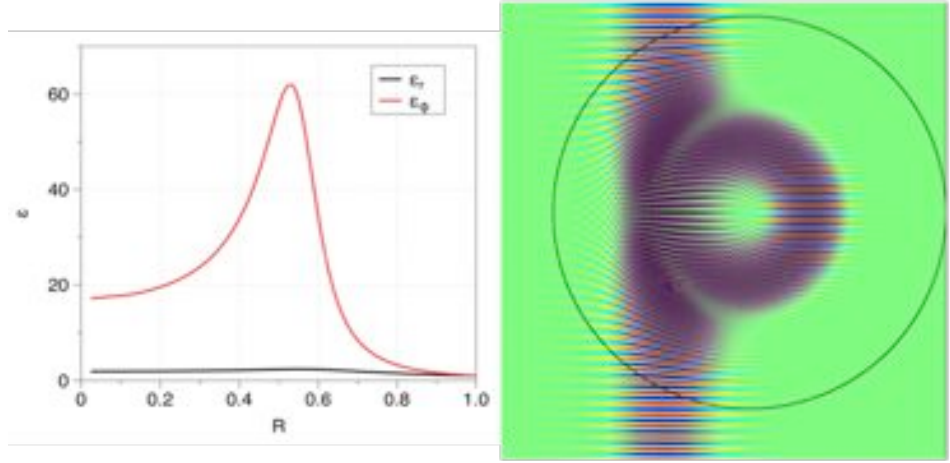

Figure 8. Left: The relevant components of the permittivity tensor for the transmuted invisible disc for light polarised with its electric field parallel to the $\{r, \varphi\}$ plane ($\mu_z=1$). Right: Full wave numerical modeling (instantaneous total electric field) of the transmuted invisible disc.

The final family of transmuted graded index devices that we will consider we will refer to as "multipole" lenses due to the fact that their ray trajectories mimic the field lines of multipolar line-charge distributions (dipolar, quadrupolar etc). The refractive index profile of such a lens is given by

$$n(r) = \frac{2r^{(2^{l-1}-1)}}{1+r^{2^l}} \qquad (14)$$

with $l = 1$ corresponding to a dipolar distribution, $l = 2$ a quadrupolar distribution etc. Examples of the ray trajectories given by eqn. 14 for $l = 0, 1$ & $2$ are shown in Fig. 9. The profile (eqn. 14) for arbitrary real $l$ is known as the generalized Maxwell fish-eye [38, 39], and for $l = 1$ this becomes the more well-known form of the Maxwell fish-eye lens in which each point on the radius of the device is refocused to the opposite point on the same surface [17]. It was noted in [40] that the ray trajectories of the fish-eye in one plane (see Fig. 9) are exactly the electric field lines of an electric dipole made up of two line charges. For integer $l > 1$ the ray trajectories similarly give exactly the electric field lines of higher-order multipoles of line charges.

Since, for all of the multipole lens profiles, the refractive index outside of the unit radius is lower than unity, it is impossible to truly mimic the ray trajectories of these multipolar modes over all space using realistic broadband materials. However, it is possible to mimic them within the unit radius by simply truncating the refractive index profile (resulting in a device of unit radius that is bounded by vacuum), and this is the approach taken here.

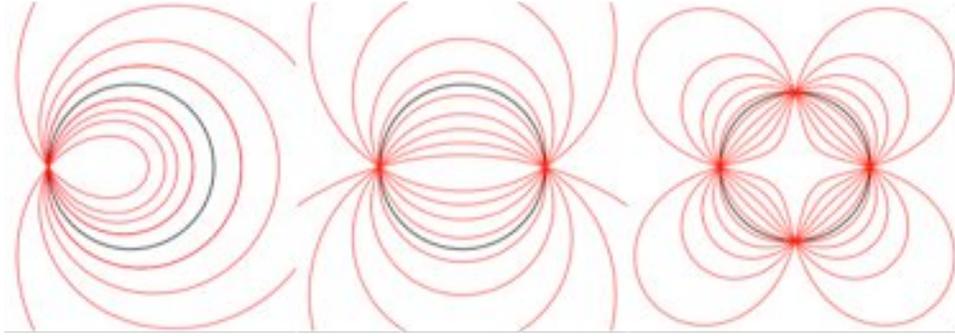

Figure 9. Ray trajectories for refractive index profiles given by eqn. 14. Left: $l = 0$, a monopolar mode, Center: $l = 1$, a dipolar mode identical to the well-known Maxwell fish-eye lens, Right: $l = 2$, a quadrupolar mode

The dipolar ($l = 1$) profile has a finite non-zero refractive index in the center of the device and as such is not of interest here. However, the $l = 0$ (monopolar) mode exhibits a singularity in the center, whilst all modes of order higher than dipolar exhibit zeros in the center. We might expect, therefore, that by using the same methods as described in section 2, we may transmute these zeros and singularities and generate practical refractive index profiles that perform the same purpose.

We will begin by considering the monopole lens; a device in which any rays emitted into the device by a point source placed at the radius are refocused back to the source. The resulting permittivity profile of the relevant tensor components after the transmutation process for light polarised with its electric field in the $\{r, \varphi\}$ plane, and the corresponding full wave modeling of the device, are shown in Fig. 10. The "bending" of the waves within the device back towards the source is clear, resulting in all of the power radiated from the source being emitted into the left hand half space.

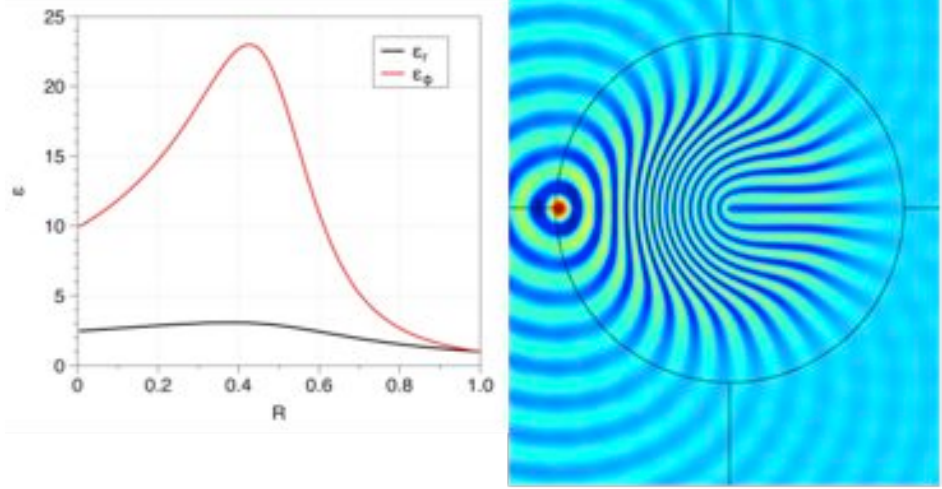

Figure 10. The resulting permittivity profile and full wave modeling (instantaneous total electric field) of the transmuted monopolar lens ($l = 0$) for light polarised with its electric field in the $\{r, \varphi\}$ plane ($\mu_z$=1). The wave components from a point source at the boundary that enter the device are re-focused back to the source.

Finally we will demonstrate the transmutation of a zero in the refractive index profile by considering the quadrupolar lens. Unfortunately, it is impossible to transmute the zero whilst maintaining all relevant permittivity components above unity when the bounding medium is vacuum. However, by simply scaling the original refractive index profile (eqn. 14) by a constant factor this can be overcome (the scaling of the refractive index does not alter the ray trajectories). In this case, this requires a scaling factor of greater than 2.4, and the resulting permittivity profile for the lens is shown in Fig. 11 ($\mu_z=1$).

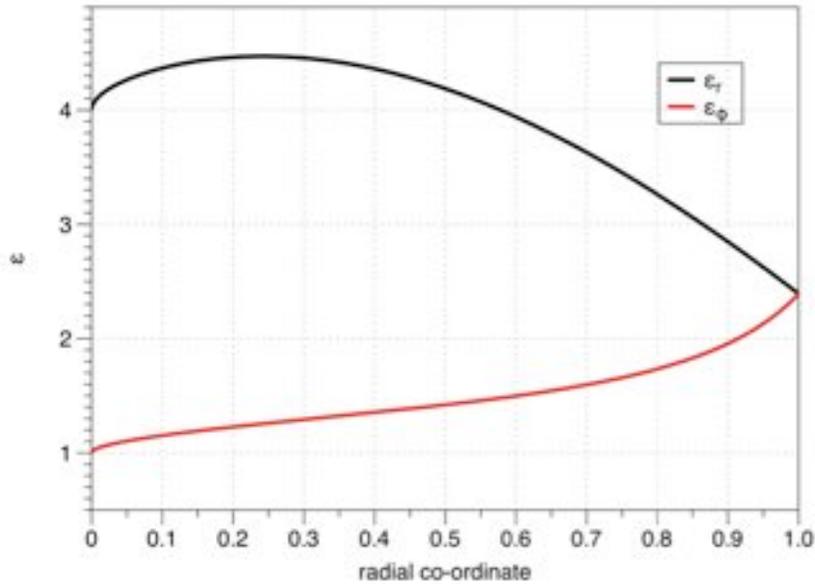

Figure 11. The permittivity profile of the transmuted quadrupolar lens for light polarised with its E-field parallel to the $\{r, \varphi\}$ plane ($\mu_z=1$). The original refractive index profile (eqn. 14) is scaled by a factor of 2.4 to enable all relevant permittivity components to be greater than unity.

The full wave modeling of the quadrupolar device for a point source placed at the radius on the left hand side is shown in Fig. 12, with the two foci evident at the top and bottom of the device demonstrating its beam-splitting character. However, it is when we place a second point source diametrically opposite the first that the quadrupolar nature of the ray trajectories within the device becomes more evident, and this is shown in the right hand panel of Fig. 12. Note that if the full refractive index profile were implemented (i.e. including the profile outside the unit radius), this second source would not be required since the rays exiting the device would loop back, completing the quadrupolar geometry of the ray trajectories shown in Fig. 9.

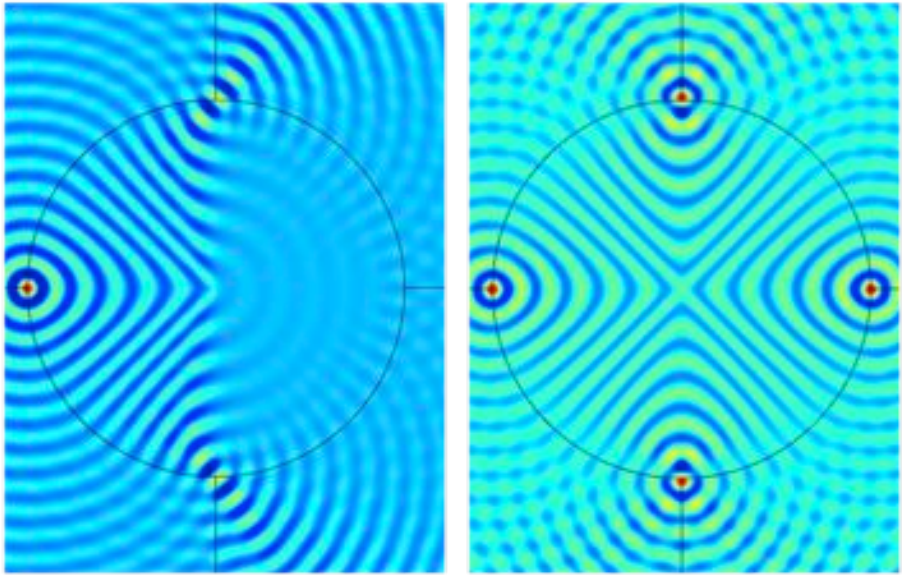

Figure 12. Full wave modeling (instantaneous total electric field) of the quadrupolar lens using the permittivity profile shown in figure 11 for light polarised with its electric field in the $\{r, \varphi\}$ plane. Left: with a single point source placed at the radius of the device on the left. Right: with two point sources, one on the radius to the left of the device, and an additional one to the right.

## 4. Conclusion

We have demonstrated the transmutation of zeros and singularities in the refractive index profiles of various optical instruments, resulting in realistic practical device designs that could in principle be fabricated using available metamaterials, and which would allow broadband low-loss operation for a single polarisation. This has been achieved using the formalism of transformation optics with the addition of a boundary condition requiring the gradient of the co-ordinate transformation be continuous. This condition ensures that there are no reflections from the boundary of the device when the resulting permittivity and permeability tensors are scaled to generate an all-dielectric design. We have demonstrated the methodology in detail using a transmuted cylindrical Eaton lens as an example, and have also shown results for an "invisible disc" and for "multipolar" lenses. However, it is important to note that the modification of the co-ordinate transformation to prevent reflections from the boundary can be applied to any transformation optics designed system. This work paves the way for transformation optics designed broadband single polarisation devices using non-resonant materials.

(All underlying research materials can be accessed by contacting I. R. Hooper).

## Acknowledgments

The authors would like to thank Prof. Roy Sambles of the University of Exeter, Dr. Robert Foster of Queen Mary, University of London, and Prof. Tomáš Tyc of Masaryk University for valuable suggestions. This work was supported by the EPSRC (UK) through the QUEST project (ref: EP/1034548/1).